# Making Effective Decisions

Machine Learning and the *Ecogame* in 1970

Catherine Mason
The Computer Arts Society
London, England
cath@catherinemason.co.uk

## ABSTRACT

This paper considers *Ecogame*, an innovative art project of 1970, whose creators believed in a positive vision of a technological future; an understanding, posited on cybernetics, of a future that could be participatory via digital means, and therefore more democratised. Using simulation and early machine learning techniques over a live network, *Ecogame* combined the power of visual art with cybernetic concepts of adaptation, feedback, and control to propose that behaviour had implications for the total system. It provides an historical precedent for contemporary AI-driven art about using AI in a more human-centred way.

## 1 Introduction

This position paper considers *Ecogame*, an early example of the convergence of art, cybernetics and computing in Britain, and asks could an interactive digital contribution to art, based on an idea of simulation put forward fifty-five years ago, make art a force for positive change using contemporary Artificial Intelligence (AI)? In 1970 *Ecogame's* creators recommended that "Man's next and most immediate task is to learn about systems and their inter-relations so that he can manage his resources and make more effective decisions." [1]

Cybernetics, driven by computing, allowed a greater degree of collaboration and engagement than was typical of art in the 1960s and 1970s. Transcending the norm, *Ecogame* offered an innovative way of considering the relationship of artist to audience, and the process of artmaking. Able to modify its reactions according to the behaviour of the participants, *Ecogame* was a simulation that modelled an economic system and became the first digitally driven, multi-player, decision-based 'game' in the UK. Developed collaboratively by members of the Computer Arts Society, *Ecogame* was part of that community's conversation around early machine learning in art and the challenges that a highly computerised society could face in the future. Its creators set out to use art to illustrate power dynamics in economic structures and their impact on ecology, and to explore potential losses of agency that they predicted could occur in a future society dominated by digitality. This historical work provides an example for contemporary AI-driven art about using AI in a responsible way – inclusive, socially relevant, and with the potential to make full use of its unprecedented promise.

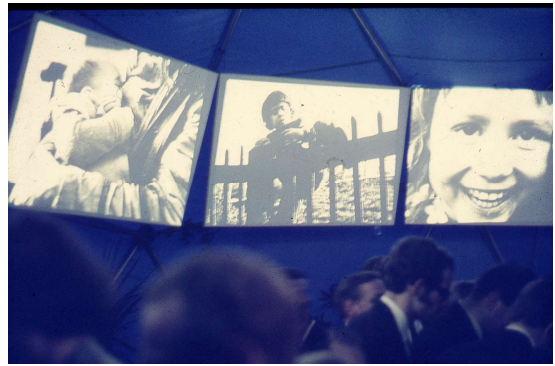

**Figure 1:** *Ecogame* presentation at Computer '70, London England. [Photograph courtesy George Mallen Archive]

## 2 The Computer Arts Society

*Ecogame*, as its name suggests, took the theme of competition and cooperation in the fields of economics and ecology. In this it tied-into the burgeoning environmental movement that began in the mid-1960s. *Ecogame* was worked on by around 25 members of the Computer Arts Society (CAS for short), for a period of ten months, led by George Mallen, one of the founders of CAS. Mallen did most of the programming at home, after business hours via a timesharing service (connected to a remote mainframe via a teletype). [2]

Digital computing technology at this time was expensive and difficult to access, being predominately located in a small number of academic, specialist research or governmental organisations, and requiring expert knowledge to operate. This then was an art form practiced by comparatively few people. For these reasons, CAS, conceived in late 1968 and officially founded in early 1969, became internationally important as a meeting place and extended community for individuals from both the technical computing side, as well as those with a fine art education. [3] Like the ground-breaking *Cybernetic Serendipity* exhibition (1968) before it, CAS emphasized the positive aspects of new technology, with an interest in exploring human-machine creative endeavours, particularly with reference to algorithmically coded interactivity. The importance of community was paramount: like-minded people able to create a space for themselves where these new ways of making art could be discussed, experimented with, exhibited, and even, occasionally, funded. In this way, computer art was both an aesthetic movement and a social movement. CAS was practitioner-led, founded by Alan Sutcliffe, a programmer, R John Lansdown, an architect, and the physicist George Mallen, and as such was able to accommodate a wide variety of approaches and processes. In fact, the name - Arts in the plural, was chosen precisely to encompass all possible art forms in which computing could play a role whether now or in the future. Everyone involved with CAS in the early days had in common an understanding of the theory of cybernetics and a desire to explore what this might mean for the arts.

## 3   Influence of Cybernetics

For these practitioners, the post-WWII science of cybernetics – the study of control and communication processes in electrical, biological and mechanical systems, [4] offered a means to consider the incorporation of disparate disciplines into art. CAS members believed this was essential in modern life, one that they saw was fast becoming computerised. Through the use of digital technology - the very basis of modern industrialised life, art would become socially relevant and positive future roles for artists ensured. This went against the norm, as most art practices of this period were commonly tied to acts of individual self-expression. However, Mallen was very open to incorporating into art all forms of knowledge including psychology, engineering and behavioural science, and introducing cooperation. In 1973 he wrote about facing the challenges of life, "No longer can we rely upon specialization of roles to give sense to the overall workings of such complexity, but we must anticipate the need to work in groups, groups which include the ability to assess scientific, social, industrial and artistic criteria." [5] Inspired by cybernetics, Mallen and colleagues considered computing as an analogy for the mind.

A particular influence was British cybernetician Gordon Pask's Conversation Theory – this had a direct impact on Mallen who had worked with Pask at his company System Research Ltd, during the mid-1960s. Conversation Theory attempts to explain learning in both living organisms and machines. This proposes that knowledge is made explicit through conversations about a subject matter and therefore an understanding of the relationships between the concepts. In 1973 Pask wrote that his Theory "emerged from experiments, perceptual motor learning, group interaction and sequential choice" developed over more than a decade. [6] Mallen and his colleagues saw that by implementing a cybernetic system in art - controlling inputs, feedback, and outputs in a framework of group interaction and sequential choice, would in turn form micro-systems within which both the artist and the participants become involved in art creation. The meaning and functionality of art could thus be extended.

## 4   Playing the Game

Delivered over a live network (connected to a remote time-sharing computer via telephone lines), *Ecogame's* hardware consisted of nine Tektronix graphics terminals (the first such terminals in Europe with tracker ball interaction), using recently invented acoustic couplers. Additionally, an Idiom minicomputer-driven interactive graphics system with large screen and light pen interaction, was also linked to the remote computer using an acoustic coupler. This was the first time such tech had been put together in this way. Participants used joysticks to input their decisions about how to deploy the investment and consumption of available resources in the model. The decisions of one player could immediately affect the resource flows to themselves and other players and ultimately the behaviour of the whole system. The results of the action were made visible in real time through projected images on screens overhead [Figure 1] and through a physical tank of water, in various states of fill. The images came from a set of 720, 35mm glass mounted slides and, under computer control, could be accessed almost instantly. This gave a visual impression of the state of the model economy. Thus, if resources were scarce, pictures were displayed showing slum housing or bad working conditions whereas, if resources were plentiful, pictures were displayed showing the happier aspects of life in the past, present, and even some projections into the future. Images were sourced from an open call to the CAS membership. Mallen described how the model was "very precisely defined [...] in which wealth is distributed through our social and industrial system, using the analogue of a reservoir of water and a plumbing system which includes a number of taps, drains and recirculating pumps." [7] That the simulation deliberately retained as much complexity and subtlety as possible, was down to the brilliance of its creators.

Each participant made decisions about how to channel the available 'wealth' in the system, what portion of that wealth should be fed back into the system, and how much kept back personally for themself. Each decision-point was presented as a four-fold choice. An example relating to the category of environment is as follows:

> "Your tanker fleet has been discharging residual oil at sea. Which option do you choose?  1: Pollute the sea with oil and employ a public relations officer to deny the fact. 2: Pay for in-port cleaning. 3: Avoid dumping within 50 miles of coastline. 4: Invest in pollution-control flushing technology." [8]

In this case, the maximum personal profit would accrue to choice 1, whilst the same choice would reap a maximum penalty in terms of social cost to the commonwealth. Questions such as these show the CAS concerns over ecology with reference to the oil industry, and the thinking around the integration of oil as a commodity in defence, health, education, and so on, and its interconnectedness to the whole economy. *Ecogame's* simulation made participants aware of the impact of parameters including personal motivation versus commonwealth, differential distribution of income, and personal sense of agency within the overall system. Through its output as a graphic, qualitative interpretation, players could clearly recognize the interconnectedness between the machinery of production and distribution, the workings of power, and associated impacts on the totality of the economic simulation in play.

## 5   Exhibition

As the interdisciplinary remit of computer art was a revolutionary concept in 1970, CAS had to find alternative exhibition possibilities. Computing manufacturers, whose primary aim was to showcase their latest technology, proved to be supportive in several instances. This led to *Ecogame* being commissioned by the BETA (Business Equipment Trade Association), to be the centrepiece of their *Computer '70* trade fair held at London's Olympia Exhibition Halls in October 1970. BETA's remit to demonstrate "one unusual, but potentially important, application of computing in a social context" [9] was specifically interpreted by CAS with a human-centred approach to show how computing in the arts could be beneficial to humankind. As early CAS member Gustav Metzger stated, to be cognoscente of the "responsibility of the artist for his material and to society." [10] This was seen as especially important given that the material in use here had initially been invented for warfare purposes.

*Ecogame* was presented under a dome, covering a space 34 feet by 43 feet, where it was played by some 5,000 people over the five days of the fair. [Figure 1]. The following year, *Ecogame* toured to Davos, Switzerland at the personal invitation of Klaus Schwab, the founder of the first European Management Symposium. Here it was played, in a slightly different format without the dome, by some of the 450 delegates from thirty-one countries at what became the first iteration of the World Economic Forum. [Figure 2]. This demonstrates the wide appeal that such a project as *Ecogame* can have, able to reach a varied audience, from laypeople to specialist management, politicians, and academic personnel.

Sadly, apart from these two exhibitions, *Ecogame* was never shown again, and its importance went unrecognised by the art world. Although the cybernetician Stafford Beer used *Ecogame's* innovative slide projection technology in his unrealised Project Cybersyn proposal for the Salvador Allende government of Chile in 1971 [11], *Ecogame's* influence ultimately remained limited to the relatively small realm of CAS. It is still little-known today.

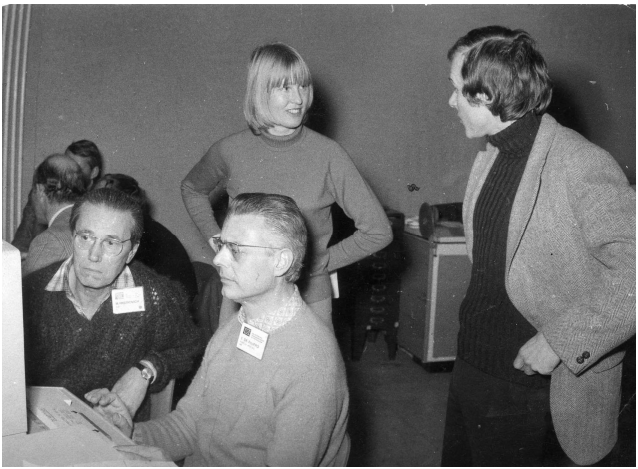

**Figure 2:** *Ecogame* **presentation at Davos, Switzerland, 1971. [Photograph courtesy George Mallen Archive]**

## 6 Machine Learning and the Role of the Artist

*Ecogame* connected to important issues of the time relating to machine learning and the role of the artist in the computational age. Would the artist of the future have any role at all, and if so, what could it look like? In a prescient paper first presented in 1970, early CAS members Ernest Edmonds and Stroud Cornock addressed this question - "Is the artist amplified or superseded by the computer?" [12] They concluded that the meaning and functionality of art could be extended by embracing interactivity and handing over some control to the audience. In this way artists would retain a prominent societal role and be amplified by computing. Today, as humanity faces climate crises and grapples with the impact of AI, many of the issues confronted by CAS seem even more pertinent.

CAS believed that a cooperative, digitally driven art based on simulation could augment human experience for the good. However, it was a radical act for an artist to position themselves between the viewer and the artwork. As Edmonds has explained, "we were all very concerned about this hierarchical view where in some people's minds the artist was some special person who dictated what was good and handed it down for the benefit for the public to get something from it. […] the artist knew everything, was a kind of god. […] we saw the good side of the computer as giving us a way out of that." [13] By handing over some of the creative experience to the viewer, the role of the artist becomes about enabling and encouraging creativity in others. In this respect, the use of new computing technologies could be used to make a more democratised art. However, in 1970 this was a profound position, completely outside the mainstream art world which tended to position an artist as a lone, creative being in the Romantic tradition. An art made with computers was therefore a subversion of established norms and CAS remained an outsider group.

## 7 The Art World and AI

As the history of a subject greatly affects how we experience that subject in our own time, an understanding of early precursor-AI projects such as *Ecogame*, can inform present conversations and debates around AI and creativity. *Ecogame* was fundamentally different from today's AI-driven image generators trained on data sets of already existing images, instigated by prompts. Instead, *Ecogame* was programmed by setting up a system, hand-coded using sets of rules, not data. As the editor of the new AI Art Magazine, featuring contemporary Generative AI work, recently commented, "[*Ecogame* was] not about the artist telling people what to think, it's about creating a system that makes players grapple with the ethical implications of their choices in a simulated world. I like that gentle way, it's a proposition. I think it's the most effective way to change the world and get into people's consciousness. I wish there was this kind of experience today more often for us as a global community to think more about the way things are going." [14] As society continues today to guide and evolve AI technology, there is potential for creative collaboration based on examples from the past.

It may be tempting to see *Ecogame* as a mere product of a more innocent, optimistic age. However there are lessons that can inform today's practice of Explainable AI in the Arts. Certainly, *Ecogame's* makers were 'closer' to their machines, which were not yet the impenetrable black boxes with highly commercialised software that we are familiar with today. By working collaboratively, they were able to hand-craft and personalise the technology, getting it to do things it was not originally designed for. In this, they committed to using it inclusively in a profoundly caring and human-centred way. Made by a community largely marginalised by the art world, *Ecogame* itself created its own interrelated community within the artwork. It aimed to spark social reflection – participants learned, in a collective manner, something about societal concerns whilst having an art experience together, via the system. As CAS co-founder John Lansdown wrote in retrospective, "It was attempting to influence people in the ways that art influences people... indirectly through the images it was creating, and the special way that an artist charges a set of images with emotion." [15] Here, the art is embedded in the behaviour and takes place in the interaction between the people involved. The flexibility of the model was one of its great strengths; both the hardware and the software could be varied to suit the implementation circumstances, making it versatile for a variety of different applications. Mallen wrote, "The environmental values were those of the *Ecogame* creators […] in different implementations, the values [could be] derived from the consensus views of others." (16).

Today, such an adaptive learning system could be utilised for a huge variety of possibilities. For example, to explain AI itself to non-specialist audiences or help users with no AI or coding expertise access these ideas through art. Such a system could create a visualisation of the power dynamics of corporate AI companies by making visible the processes involved in decision-making, and to spark personal reflection on the ethical implications of one's choices. The current heated debates over copyright and to what extent who stole whose material without permission is an unnecessary smokescreen detracting from much larger and urgent debates. In the mad rush of AI image generators that churn out 1000s of images, it is worth reflecting on a practice that is more mindful of how we can use this technology to understand and even go some way towards mitigating the negative effects of the digitality that we now find ourselves completely immersed in.

Historic projects such as *Ecogame* remind us that digital tools reach their fullest potential when used to build community, amplify human capability and inspire change through creativity, demonstrating ways in which we are all inter-connected. The early computer artists pushed the technology of their day to its absolute limits, to create an art that not only spoke about their social challenges, but dealt with them in a didactive, creative and above all accessible manner. In this they were not ahead of their time, rather they were very much OF their time.

The example of CAS and *Ecogame* show that to truly make art a force for positive change today, communities consisting of designers, coders, technology developers and artists need to be encouraged to work together to involve audiences in art interactions. This will assist with increased usability and transparency of AI-infused experiences. We need to use creative computing to empower people to find ways of making projects that can channel the never-ending flow of information in our lives into curated participatory experiences. This undoubtedly will require difficult conversations, for example, about proper remuneration of creative labour, and the need to regulate the consumerism of Big Tech, among other things. However, in this way there will continue to be a crucial role for both AI and artists in storytelling and enabling audiences, via art, to create their own stories.

## REFERENCES

[1] Foundation General Systems Ltd: John McNulty, Christine McNulty, Anthony McCall. 1970. Statement. In *PAGE*, no. 16, June. n.p.

[2] Catherine Mason (Ed.). 2024. *Creative Simulations: George Mallen and the Early Computer Arts Society*. Springer, Cham, Switzerland.

[3] Catherine Mason, 2008. *A Computer in the Art Room: the Origins of British Computer Arts 1950-1980*. JJG, Norfolk, UK.

[4] Norbert Wiener, 1948. *Cybernetics: Or Control and Communication in the Animal and the Machine*. Hermann & Cie, Éditeurs; The Technology Press, Cambridge; John Wiley & Sons Inc. New York.

[5] George Mallen. 1973. Wherever Next. In *Computers in the Creative Arts (A Studyguide)*. Jeffrey David Lomax (Ed.). National Computing Centre Publications, Manchester, UK. 55-58.

[6] Gordon Pask. 1976. *Conversation Theory: Applications in Education and Epistemology*, Elsevier Amsterdam, Oxford, New York, preface ix.

[7] George Mallen. 1971. Towards the Concept of an Art System. In *Proceedings of Symposium Creativity in a Machine Environment*. Stroud Cornock (Ed.). City of Leicester Polytechnic, Leicester, UK. n.p.

[8] George Mallen. 2024. From Cybernetics to *Ecogame*: Computing in a Cultural Context - An Interview with George Mallen. In: Catherine Mason. (Ed.) *Creative Simulations: George Mallen and the Early Computer Arts Society*. Springer Cham, Switzerland. 45-67.

[9] Computer Arts Society. 1970. *PAGE,* no. 8, May. n.p.

[10] Gustav Metzger. 1969. Automata in History. In *Studio International*, 3. 107.

[11] Catherine Mason. (Ed.) *Creative Simulations: George Mallen and the Early Computer Arts Society*. Springer Cham, Switzerland. 65.

[12] Stroud Cornock and Ernest Edmonds. 1973. The Creative Process Where the Artist Is Amplified or Superseded by the Computer. In *Leonardo*, vol. 6, no. 1. 11-16.

[13] Ernest Edmonds. 2022. Ernest Edmonds in Conversation with Catherine Mason. In *Ernest Edmonds: Art:Notes +Works*. Boco Publishing, Sheffield, UK. 13-17.

[14] Simone Brauner. 2024. Conversation with author 11 October 2024.

[15] John R. Lansdown. 1984. Interviews. In Brian Reffin Smith. *Soft Computing: Art & Design*. Addison-Wesley, Wokingham, UK. 147-178.

[16] As note 5.